\documentclass[letter]{aa} 
\usepackage{graphicx}
\usepackage{txfonts}
\usepackage{hyperref}
\graphicspath{{Images/}{../Images/}}
\usepackage{xcolor}

\begin{document}

    \title{Giant exoplanet composition:
    }
    \subtitle{Why do the hydrogen-helium equation of state and interior structure matter?}

    \author{S. Howard \inst{1}
          \and R. Helled\inst{1}
          \and S. Müller\inst{1}
          }

    \institute{Institut für Astrophysik, Universität Zürich, Winterthurerstr. 190, CH8057 Zurich, Switzerland,\\
              \email{saburo.howard@uzh.ch}
              }
    \date{}
    
 
  \abstract
   {Revealing the internal composition and structure of giant planets is fundamental for understanding planetary formation. 
   However, the bulk composition can only be inferred through interior models. As a result, advancements in modelling aspects are essential to better characterise the interiors of giant planets.
   }
   {We investigate the effects of model assumptions such as the interior structure and the hydrogen-helium (H-He) equation of state (EOS) on the inferred interiors of giant exoplanets.
   }
   {We first assess these effects on a few test cases and compare H-He EOSs. We then calculate evolution models and infer the planetary bulk metallicity of 45 warm exoplanets, ranging from 0.1 to 10~$M_{\rm J}$.
   }
   {Planets with masses between about 0.2 and 0.6~$M_{\rm J}$ are most sensitive to the H-He EOS.
   Updating the H-He EOS reduces the inferred heavy-element mass, with an absolute difference in  bulk metallicity of up to 13\%. 
   Concentrating heavy elements in a core, rather than distributing them uniformly (and scaling opacities with metallicity), reduces the inferred metallicity (up to 17\%). The assumed internal structure, along with its effect on the envelope opacity, has the greatest effect on the inferred composition of massive planets ($M_{\rm p}>4~M_{\rm J}$). 
   For $M_{\rm p}>0.6~M_{\rm J}$, the observational uncertainties on radii and ages lead to uncertainties in the inferred metallicity (up to 31\%) which are larger than the ones associated with the used H-He EOS and the assumed interior structure. However, for planets with $0.2<M_{\rm p}<0.6~M_{\rm J}$, the theoretical uncertainties are larger.
   }
   {Advancements in equations of state and our understanding of giant planet interior structures combined with accurate measurements of the planetary radius and age are crucial for characterising giant exoplanets. 
   }

   \keywords{planets and satellites: gaseous planets --
                planets and satellites: interiors --
                planets and satellites: composition
               }

   \maketitle
%
\section{Introduction}

Determining the bulk composition and internal structure of giant planets is fundamental for understanding their origin \citep[e.g.,][]{mordasini2016,turrini2018}. Observations of giant exoplanets have advanced significantly and nowadays, not only the planetary mass and radius are measured, but also the atmospheric composition \citep[e.g.,][]{kreidberg2014,madhusudhan2014,edwards2023}. However, the bulk composition of giant exoplanets is typically inferred through models, where the planetary structure is constrained by mass, radius and age measurements. \\
\indent Accurate measurements of the gravitational fields and atmospheric composition of Jupiter and Saturn have led to the development of a more comprehensive theoretical framework for giant planet modelling. 
Interior models of both Jupiter and Saturn that fit Juno \citep{bolton2017} and Cassini \citep{spilker2019} data suggest that the planets have dilute cores and complex interiors \citep{wahl2017,mankovich2021,miguel2022,howard2023_interior,helled2024,howard_invZ,muller2024}. In addition,  giant planets models strongly depend on our understanding of the hydrogen-helium equation-of-state (H-He EOS) \citep[e.g.,][]{militzer2013,chabrier2019,howardguillot2023}.\\
\indent As observational uncertainties are progressively being reduced, thanks to ground-based observations (e.g., HARPS \citep{mayor2003}, NIRPS \citep{bouchy2017}, ESPRESSO \citep{pepe2021}) and space missions (e.g., Kepler \citep{borucki2010}, TESS \citep{ricker2015}, JWST \citep{gardner2006} and in the future PLATO \citep{rauer2014}, Ariel \citep{tinetti2018}), it is clear that more advanced models should be used for the characterisation of giant exoplanets. In addition, it is important to identify the theoretical uncertainties and compare them to the observational ones. \\
\indent Several previous studies have explored the importance of model assumptions such as the interior structure and the EOS on the inferred planetary bulk composition \citep{baraffe2008,muller2020,bloot2023}. In addition, the large sample of observed giant exoplanets with measured radius, mass and age, together with theoretical models allows inferring the relationship between the planetary mass and its metallicity \citep{thorngren2016,muller2023}. 
The goal of this work is to further investigate the effects of model assumptions on the inferred internal structures of giant exoplanets. Our models use the most up-to-date H-He EOS, allow for Core+envelope and Fully-mixed interiors, and the opacity scales with the metallicity. We improve upon previous work by systematically studying how the H-He EOS and the distribution of heavy elements affect the inferred bulk composition and mass-metallicity relations.\\
\indent Our paper is organised as follows. Section \ref{sec:methods} describes our methods. In Sec.~\ref{sec:testcases} we analyse the effects of the assumed interior structure and hydrogen-helium (H-He) EOS on a few test cases. Section \ref{sec:planetS} focuses on a sample of observed exoplanets. Our conclusions are presented in Sec.~\ref{sec:conclusion}. 

\section{Methods}
\label{sec:methods}

We used CEPAM \citep{guillot1995_cepam, howard2024} to run planetary evolution models and assess the effects of both the H-He equation-of-state (EOS) and the assumed interior structure (i.e., the distribution of heavy elements). We compared calculations using the SCvH95 \citep{saumon1995} and the CMS19+HG23 EOSs\citep{chabrier2019,howardguillot2023}. More details about the differences between these EOSs are presented in Sec.~\ref{sec:testcases}. We also compared models assuming either a "Core+envelope" or "Fully-mixed" structure. In the first structure, all the heavy elements were concentrated in a central core. We used the analytical EOS of \citet{hubbard1989} to calculate the pressure-dependent density in the isothermal core, which was assumed to be made of 50\% ices and 50\% rocks. In the second structure, the heavy elements were uniformly distributed within the planet. We also assumed an ice-to-rock ratio of unity and used the SESAME water EOS for ices and the SESAME drysand EOS for rocks \citep{sesame1992}. The hydrogen-to-helium ratio in the envelope was protosolar: $Y/(X+Y)=0.27$ \citep{asplund2021}. Our models used a non-gray atmosphere \citep{parmentier2015} and the method from \citet{valencia1013} to account for the opacity enhancement due to the heavy elements.

\section{Proof of concept: metallicities of synthetic giant planets}
\label{sec:testcases}

We first calculated models of synthetic planets, with masses of 0.3, 1 and 3~$M_{\rm J}$. We focused on planets that are not highly irradiated and hence not inflated \citep[e.g.,][]{thorngren2018,fortney2021}, with an equilibrium temperature $T_{\rm eq}$ of 500~K. For each planetary mass, we ran models with both H-He EOSs (SCvH95, CMS19+HG23) and internal structures (Core+envelope, Fully-mixed). We assumed different bulk metallicities $Z$, from 0 to 0.5 ($Z=M_{Z}/M_{\rm p}$ where $M_{Z}$ and $M_{\rm p}$ are the total heavy-element mass and the planet's mass, respectively). 

\begin{figure}[h]
   \centering
   \includegraphics[width=0.85\hsize]{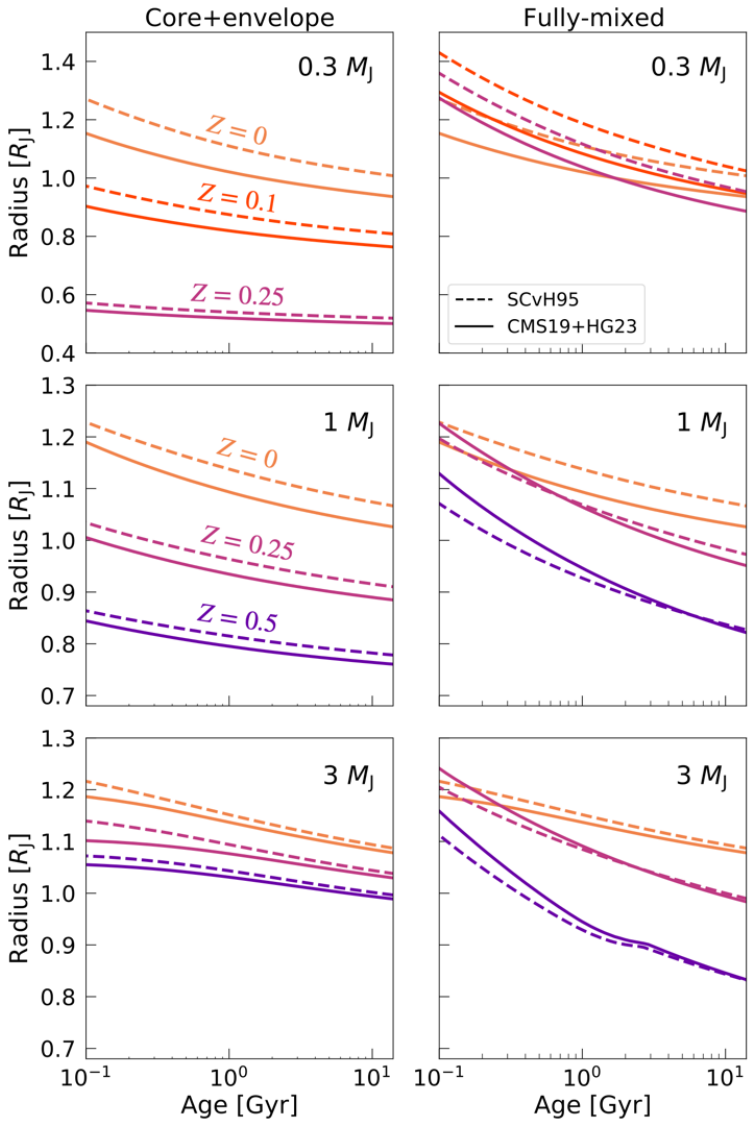}
      \caption{Radius vs.~time for planets with masses of 0.3, 1 and 3~$M_{\rm J}$ (top, middle, and bottom panels, respectively). Left (right) panels correspond to models assuming a Core+envelope (Fully-mixed) structure. Results with different H-He EOSs are shown with dashed and solid lines. The colors correspond to different assumed metallicities. The middle and bottom panels have similar limits on the y-axis, but not the top panels.
     }
         \label{figure:testcases}
\end{figure}

Figure~\ref{figure:testcases} shows the planetary radius vs.~time for the different cases. For the Core+envelope structure, CMS19+HG23 always yields smaller radii than SCvH95, by up to 10~\% (\citet{muller2020} found that CMS19 also yields smaller radii). We also find that higher metallicities lead to smaller radii. The range of radii spanned by models with different metallicities is larger for lower-mass planets. 
For $M_{\rm p}=0.3~M_{\rm J}$, the radius decreases by about a factor of two when changing $Z$ from 0 to 0.25,  while for $M_{\rm p}=3~M_{\rm J}$, changing $Z$ from 0 to 0.5 changes the radius by only 15\%. In the Core+envelope case,  for a given measured radius and its associated uncertainty, the range of inferred metallicity is narrower for a 0.3~$M_{\rm J}$ planet compared to a 3~$M_{\rm J}$ planet. In addition, for a given measured radius and its uncertainty, the Core+envelope structure leads to a narrower range of inferred metallicity for a 0.3~$M_{\rm J}$ planet, compared to a Fully-mixed structure. However, the opposite behaviour is expected for a 3~$M_{\rm J}$ planet. The comparison between both structures depends on the planetary mass and metallicity.

In the Fully-mixed case, higher metallicities do not always lead to smaller radii. This is because of two opposite effects due to enriching the envelope with heavy elements: it increases both the mean molecular weight and the opacities \citep{guillot2005}. At early times, the opacity effect seems to dominate and delays the cooling and contraction of the planet; but at late times, the mean molecular weight effect wins. \citet{muller2020} also found that the radius is not monotonically decreasing with $Z$ if the effect of the heavy elements on the opacity is accounted for. For $M_{\rm p}=0.3~M_{\rm J}$, we find that CMS19+HG23 always yields smaller radii than SCvH95. However, this is not the case for $M_{\rm p}=1$ and 3~$~M_{\rm J}$. For these two planetary masses, especially at early times, CMS19+HG23 yields larger radii than SCvH95 for $Z>0$. 

Understanding these results requires investigating the differences between the SCvH95 and CMS19+HG23 EOSs. 
SCvH95 is based on the so-called chemical picture \citep[see][for a detailed description]{saumon1995}. It models the interactions between molecules or atoms by using pair-potentials. The Helmholtz free energy is determined through the free-energy minimisation technique and pressure, entropy, or internal energy can then be calculated \citep{hummer1988}. However, at densities corresponding to dissociation and pressure ionisation, pair-potentials fail at correctly describing describing the behaviour of matter. Using the additive volume law, one can obtain the properties (density and entropy at given pressures and temperatures) of a H-He mixture. On the other hand, CMS19+HG23 also applies the additive volume law to combine pure H and He tables from \citet{chabrier2019}, while additionally accounting for non-ideal mixing effects as described by \citet{howardguillot2023}. It accounts for different EOS sources for different regions of the H-He phase diagram using in particular the data from \citet{militzer2013}. Based on the physical picture, these \textit{ab initio} calculations focus on electrons and nuclei and model their interactions by Coulomb potentials. \citet{militzer2013} considered a H-He mixture (in proportions close to protosolar) and provided a more realistic description of the regime of dissociation and ionisation. 

\begin{figure}[h]
   \centering
   \includegraphics[width=0.85\hsize]{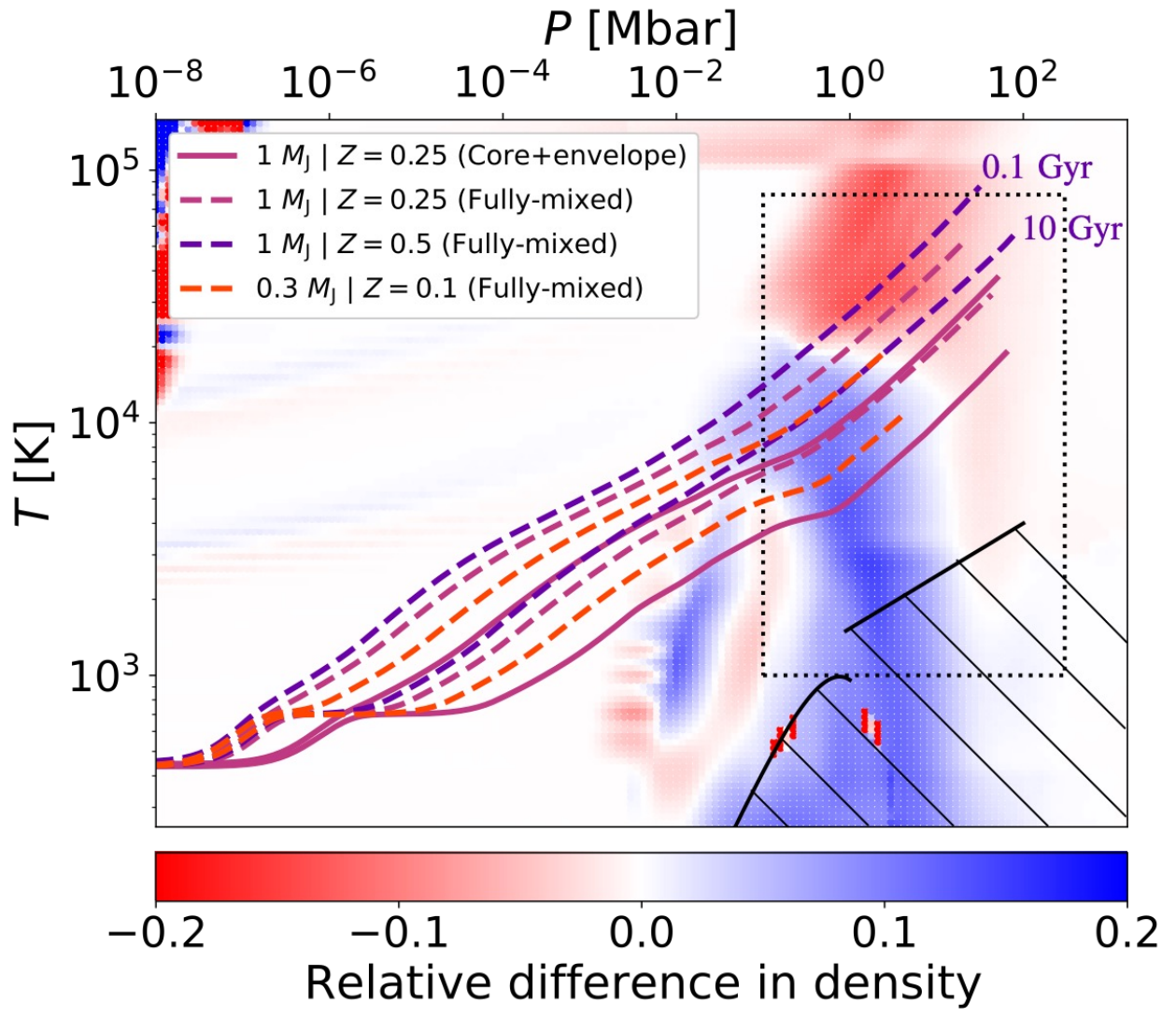}
      \caption{Relative difference in density between the CMS19+HG23 \citep{chabrier2019,howardguillot2023} and SCvH95 \citep{saumon1995} EOSs. Blue regions on the $T$-$P$ diagram indicate where CMS19+HG23 is denser than SCvH95, while red regions highlight less dense areas. The coloured solid and dashed lines show $T$-$P$ profiles of some models (using CMS19+HG23) from Fig.~\ref{figure:testcases}, at 0.1 or 10~Gyr. The dotted square indicates the region of the \textit{ab initio} presented by \citet{militzer2013}. The  CMS19+HG23 EOS is more accurate in this region. The hashed area corresponds to the region where hydrogen becomes solid \citep[see e.g.,][]{chabrier2019}.
      }
         \label{figure:hg23_vs_scvh}
\end{figure}

Figure~\ref{figure:hg23_vs_scvh} shows the relative difference in density between CMS19+HG23 and SCvH95. We include the $T$-$P$ profiles of some models from Fig.~\ref{figure:testcases}, at 0.1 and 10~Gyr. 
We calculate the differences assuming a H-He mixture of protosolar composition (He mass fraction of $Y=0.27$), and interpolate the density from CMS19+HG23 at the pressure and temperature points corresponding to SCvH95. 
In the region where {\it ab initio} calculations from \citet{militzer2013} have been performed (dotted black square), the $T$-$P$ profiles of our models can go through regions where CMS19+HG23 is denser than SCvH95 (shown in blue) or less dense (shown in red). We note that entropy influences a planet's position on the $T$-$P$ diagram by affecting the calculation of the adiabatic gradient. The density differences between the EOSs under these  $T$-$P$ conditions then explain the differences in inferred planetary radii.

We find that $1~M_{\rm J}$ with $Z=0.25$ and a Core+envelope structure (solid magenta line) mostly covers the region where CMS19+HG23 is denser than SCvH95 (around 1~Mbar). Interestingly, we find that the $T$-$P$ profiles of models assuming the Core+envelope structure are rather similar even the planetary mass or metallicity are changed. This explains why we always found that CMS19+HG23 yielded smaller radii compared to SCvH95 in the Core+envelope case (see left panels of Fig.~\ref{figure:testcases}). However, $T$-$P$ profiles of Fully-mixed models are more affected by changing the mass and metallicity. These models have a much hotter interior because enhancing the opacities due to heavy-element enrichment in the envelope delays the planetary cooling. The $0.3~M_{\rm J}$ model shown in Fig.~\ref{figure:hg23_vs_scvh} (dashed red line) is still in a $T$-$P$ regime where CMS19+HG23 is denser than SCvH95, explaining why CMS19+HG23 yielded smaller radii than SCvH95 for $M_{\rm p}=0.3~M_{\rm J}$ in the Fully-mixed case (see top right panel of Fig.~\ref{figure:testcases}). The $1~M_{\rm J}$ model with $Z=0.25$ (dashed magenta line) goes through the less dense region at 0.1~Gyr but is then mostly affected by the denser region at 10~Gyr. This explains why CMS19+HG23 yielded larger radii at early times and smaller radii at late times (middle right panel of Fig.~\ref{figure:testcases}). Overall, the differences in radii between CMS19+HG23 and SCvH95 seen in Fig.~\ref{figure:testcases}, depend on the position of the planet within the temperature-pressure-density space during its evolution. As CMS19+HG23 models the regime of dissociation and ionisation more accurately, this H-He EOS should be used for exoplanet characterisation.

\section{Inferred metallicities of observed giant  exoplanets}
\label{sec:planetS}

In this section we characterised a sample of observed giant exoplanets. We used the PlanetS catalog \citep{otegi2020,parc2024} and selected planets with masses from 0.1 to 10~$M_{\rm J}$ with relative measurement uncertainties smaller than 10 and 8\% for the mass and radius, respectively. We selected planets with $T_{\rm eq}<1000~$K and with an age estimate. We ended up with a sample of 45 exoplanets that allowed a reliable determination of the internal composition. The planets in our sample are listed in Appendix~\ref{app:A}.

\begin{figure*}[h]
   \centering
   \includegraphics[width=0.73\hsize]{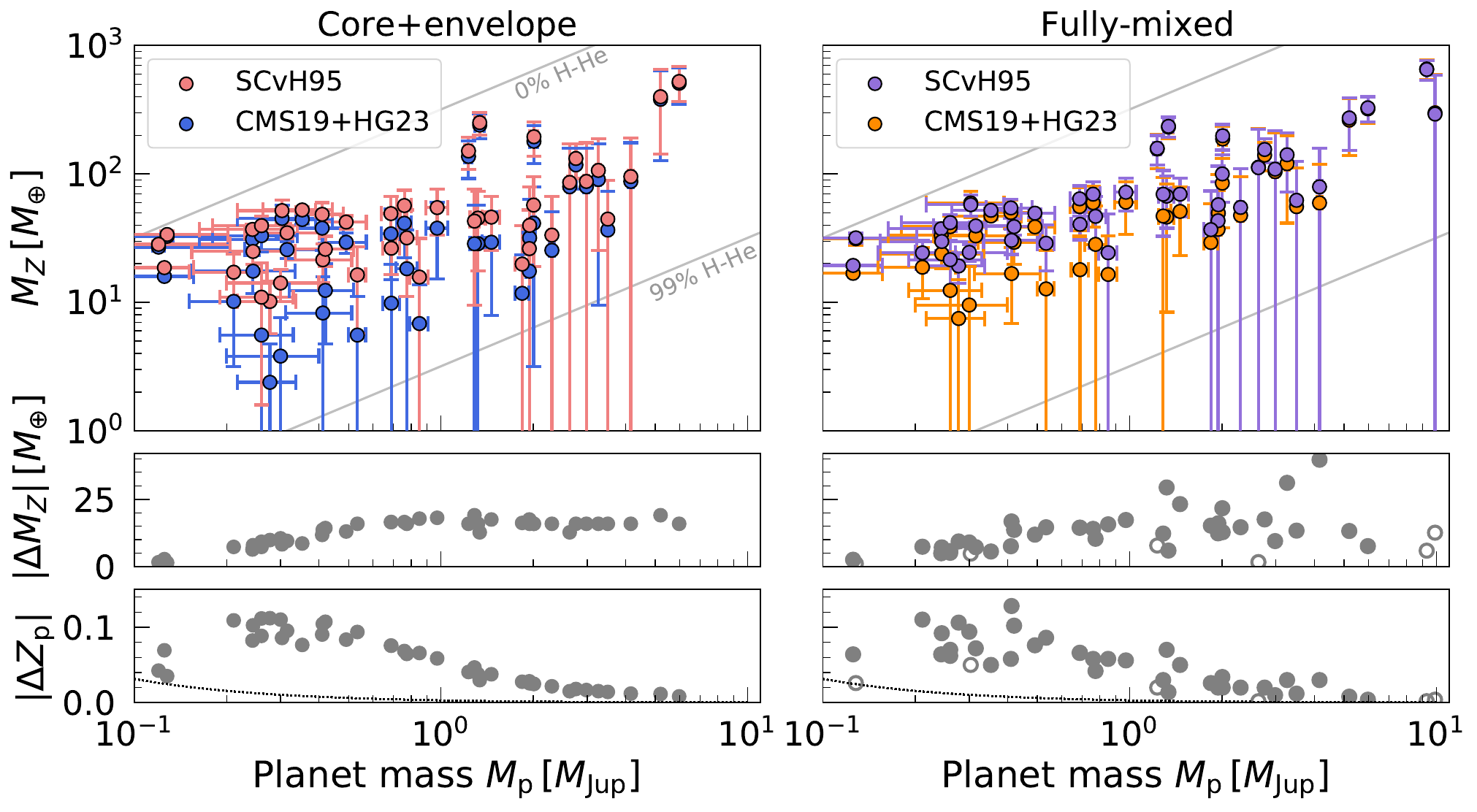}
      \caption{Inferred heavy-element mass vs.~planetary mass. {\it Top panels}: comparison of the results obtained with CMS19+HG23 and SCvH95, for both Core+envelope and Fully-mixed structures. The circles represent the midpoint between the upper and lower bounds of the errorbars. {\it Middle panels}: absolute difference in heavy-element mass due to the H-He EOS update. {\it Bottom panels}: absolute difference in bulk metallicity. Empty circles correspond to planets for which CMS19+HG23 leads to a higher metallicity than SCvH95. The black dotted line shows the $1/M_{\rm p}$ curve.}
         \label{figure:planetS_eos}
\end{figure*}

For each exoplanet, we inferred the heavy-element mass ($M_{Z}$) by calculating two evolution models which fit the upper and lower bounds of the radius and age measurements ($R_{\rm obs}^{\rm min},\rm age_{\rm obs}^{\rm min}$ and $R_{\rm obs}^{\rm max},\rm age_{\rm obs}^{\rm max}$, listed in Table~\ref{tab:param}). This provided the possible range of the inferred bulk metallicity ($Z_{\rm p}=M_{Z}/M_{\rm p}$). The inferred heavy-element mass for our sample of exoplanets is shown in Fig.~\ref{figure:planetS_eos} (top panels). We also show the absolute difference in the inferred heavy-element mass ($\Delta M_{Z}$) between both EOSs (middle panels) as well as the absolute difference in the bulk metallicity ($\Delta Z_{\rm p}=\Delta M_{Z}/M_{\rm p}$) (bottom panels). 

For the Core+envelope case, we find that: (i) CMS19+HG23 yields a lower heavy-element mass than SCvH95 for all planets. (ii) $\Delta M_{Z}$ increases with $M_{\rm p}$ in the range from $M_{\rm p}=0.1~M_{\rm J}$ to $M_{\rm p}=1~M_{\rm J}$, after which it reaches a plateau at $\Delta M_{Z} \sim 15~M_{\oplus}$ for planets with masses larger than $1~M_{\rm J}$. (iii) $\Delta Z_{\rm p}$ peaks at $M_{\rm p}=0.3~M_{\rm J}$, with a value of $\sim$11\%. In the Fully-mixed case, we find that: (i) CMS19+HG23 also yields a lower heavy-element mass than SCvH95 except for six planets. This includes the two most massive planets from our sample, as expected (see Sec.~\ref{sec:testcases}). (ii) The variation of $\Delta M_{Z}$ is similar to  the Core+envelope case. However, there is greater scatter from $M_{\rm p}>1~M_{\rm J}$. (iii) $\Delta Z_{\rm p}$ peaks at $M_{\rm p}=0.4~M_{\rm J}$, with a maximum value of about 13\%.

We find that the inferred bulk metallicity of planets with masses between $0.2$ and $0.6~M_{\rm J}$ is most affected by the H-He EOS update. This holds for both interior structures and the absolute difference in bulk metallicity can go up to 13\%. This is due to how the planets lie in the temperature-pressure-density space (Fig.~\ref{figure:hg23_vs_scvh}).
Planets with $M_{\rm p} < 0.2~M_{\rm J}$ are less affected by changing the H-He EOS as they only have a small amount of H-He. We also note that CMS19+HG23 leads to eight additional planets (compared to SCvH95) for which the lower limit of the inferred heavy element mass is zero (in Table~\ref{tab:param}, planets with a single dagger). For these planets, the radius inferred from a pure H-He model cannot match the upper bound of the measured radius suggesting that these planets may be inflated.

We next compare the heavy-element content obtained with the Core+envelope and Fully-mixed structures, using CMS19+HG23 since it is the better EOS for modelling planets. This time, we show the mass-metallicity relation by calculating the ratio of the planet metallicity $Z_{\rm planet}$ to that of the parent star $Z_{\rm star}$. We follow \citet{muller2023} and perform a Bayesian linear regression. The parameters $\alpha$ and $\beta$ of the power-law $Z_{\rm planet}/Z_{\rm star}=\beta \times M[M_{\rm J}]^{\alpha}$ are estimated using a Markov chain Monte Carlo (MCMC) method and their posterior distributions are shown in Appendix~\ref{app:C}. Figure~\ref{figure:fm_vs_ce} (top panel) shows for both structures the best fit as well as the $1 \sigma$ error contour. 

In the Fully-mixed case, we find $\alpha=-0.37 \pm 0.07$ and $\beta=8.61 \pm 0.88$. The mass-metallicity relation we obtain with the Fully-mixed structure is in line (within $1 \sigma$) with \citet{muller2023} who found $\alpha=-0.37 \pm 0.14$ and $\beta=7.85 \pm 1.49$, using CMS19. They considered opacity scaling with metallicity (with a different method than this work) and assumed a core with an enriched envelope for planets below $5~M_{\rm J}$, while adopting a Fully-mixed structure for those exceeding $5~M_{\rm J}$. The relation is also in rather good agreement with \citet{thorngren2016} who found $\alpha=-0.45 \pm 0.09$ and $\beta=9.70 \pm 1.28$, using SCvH95. They did not account for opacity enhancement and adopted a hybrid approach with a $10~M_{\oplus}$ core and the rest of the heavy elements in the envelope. However, the mass-metallicity relation we obtain with the Core+envelope structure is steeper. We find $\alpha=-0.57 \pm 0.13$ and $\beta=5.09 \pm 0.95$. The Core+envelope structure yields lower metallicities for 36 out of 42 planets (blue squares, middle panel of Fig.~\ref{figure:fm_vs_ce}). For the six other planets, the Fully-mixed structure yields lower metallicities (orange squares), and especially for massive planets ($M_{\rm p}>4~M_{\rm J}$). Fully-mixed models require more heavy elements, as their contraction is delayed by the opacity enhancement resulting from the enriched envelope (see Sec.~\ref{sec:testcases}). However, the lack of planets with high masses ($M_{\rm p}>5~M_{\rm J}$) and the inability to find solutions for the two heaviest planets under the Core+envelope structure may impact the results and contribute to a steeper mass-metallicity relation. Furthermore, the Fully-mixed case is expected to be more representative of massive planets \citep{thorngren2016,muller2021}. Further discussion about the mass-metallicity relation is given in Appendix~\ref{app:D}.

We next focus on the absolute difference in bulk metallicity $\Delta Z_{\rm p}$ for different model assumptions. We find that $\Delta Z_{\rm p}$ can go up to 17\% due to the assumed interior structure and go up to 13\% due to the H-He EOS update (middle panel of Fig.~\ref{figure:fm_vs_ce}). Overall, the interior structure and the H-He EOS seem to have comparable effects on the inferred planetary metallicity. Only for $M_{\rm p}>4~M_{\rm J}$, the interior structure has a significantly larger effect on the inferred metallicity than the H-He EOS. We then sum the effects of both the H-He EOS and the interior structure (pink dots, bottom panel) and compare it to the inferred range of metallicity (black triangles) that arises from the uncertainty in the measured radii and ages. This uncertainty range corresponds to the length of the errorbars shown on the top panel and is defined as $Z_{\rm p}(R_{\rm obs}^{\rm min},\rm age_{\rm obs}^{\rm min})-Z_{\rm p}(\it{R}_{\rm obs}^{\rm max},\rm age_{\rm obs}^{\rm max})$. It can reach up to 31\%. 
Currently, for most planets with $M_{\rm p}>0.6~M_{\rm J}$, the uncertainties on radii and ages lead to uncertainties in the inferred metallicity larger than the ones associated with the used H-He EOS and the assumed interior structure. However, for planets with masses between $0.2$ and $0.6~M_{\rm J}$ the theoretical uncertainties are larger. As a result, great caution should be taken when characterizing intermediate-mass giant planets. In addition, as future measurements are expected to have smaller uncertainties, the details of the theoretical modelling should be considered for all giant planets. 

\begin{figure}[h]
   \centering
   \includegraphics[width=\hsize]{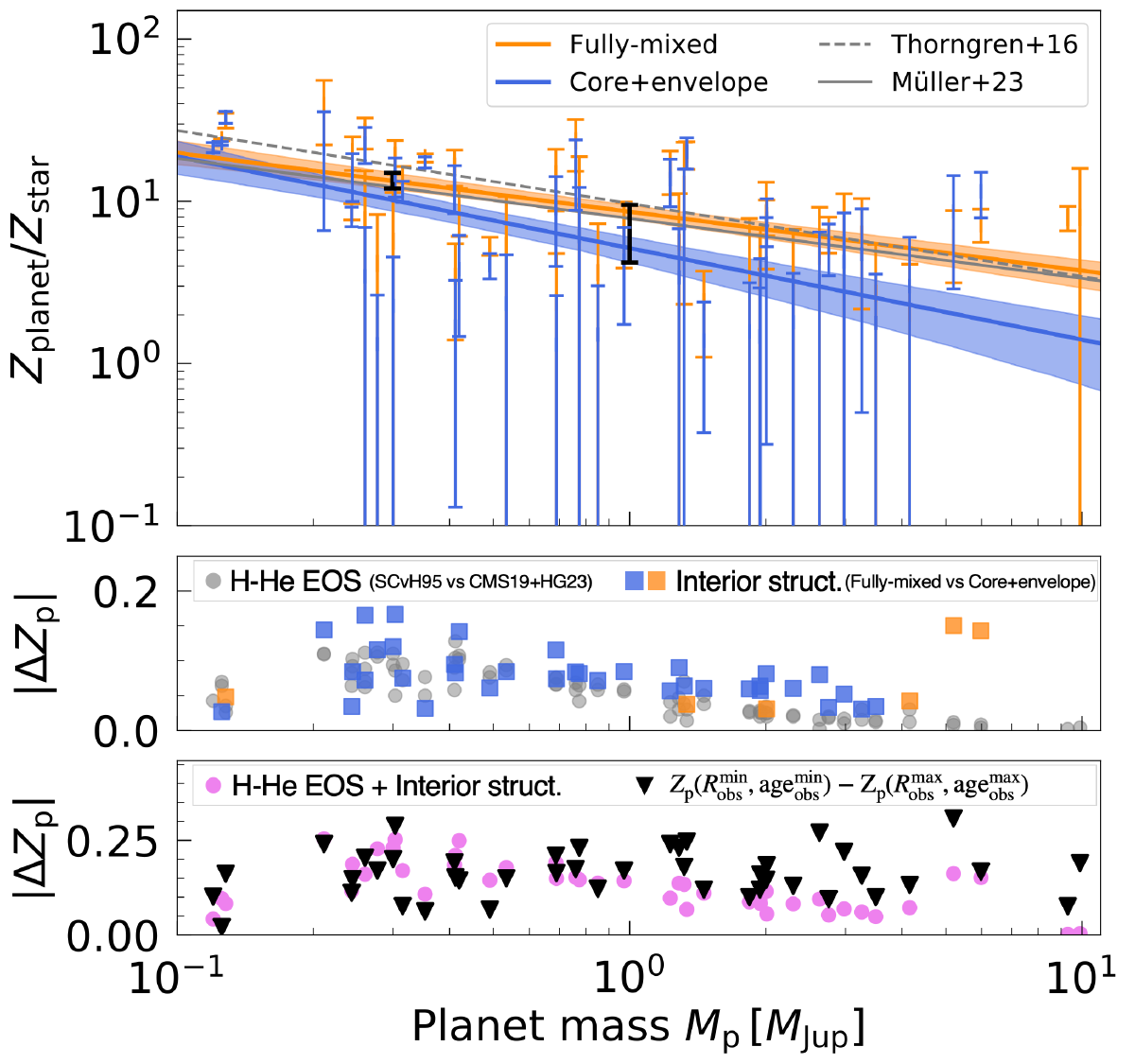}
      \caption{Mass-metallicity relation with both interior structures, using the CMS19+HG23 EOS. \textit{Top panel}: errorbars show the inferred range of bulk metallicity with the Core+envelope (blue) or Fully-mixed (orange) structure, as in Fig.~\ref{figure:planetS_eos}. Orange and blue solid lines show the best fit for both cases. The shaded areas show the $1\sigma$ error contour around the best fit. The metallicities of Jupiter and Saturn \citep{helledhoward2024} are shown in black.
      The solid and dashed gray lines show the best fits from \citet{muller2023} and \citet{thorngren2016}, respectively. The star metallicities have been calculated using the solar value from \citet{asplund2021}. \textit{Middle panel}: absolute difference in metallicity due to either the H-He EOS (gray dots, taken from Fig.~\ref{figure:planetS_eos}) or the interior structure (squares). Blue (orange) squares correspond to planets for which the Core+envelope case leads to a lower (higher) metallicity than the Fully-mixed case. \textit{Bottom panel}: absolute difference in metallicity. Pink dots show the sum of the H-He EOS and the interior structure effects. Black triangles show the range of metallicity inferred from observational uncertainties (radius and age).
      }
         \label{figure:fm_vs_ce}
\end{figure}

\section{Conclusions}
\label{sec:conclusion}

Using  a sample of 45 warm giant exoplanets, ranging from 0.1 to 10~$M_{\rm J}$, we investigated the effects of the used H-He EOS and assumed internal structure on the inferred planetary bulk composition. These assumptions are important because there is an interplay between the internal structure which defines the planet's $T$-$P$ profile and its thermodynamical properties (e.g., density, entropy) that are calculated from the EOSs (Fig.~\ref{figure:hg23_vs_scvh}).  
Our main conclusions are:

\begin{enumerate}
    \item Planets with $0.2<M_{\rm p}<0.6~M_{\rm J}$ are most sensitive to the H-He EOS. Using the CMS19+HG23 EOS rather than SCvH95 reduces the inferred planetary bulk metallicity. The corresponding absolute difference can reach up to 13\%.
    \item Assuming a Core+envelope rather than a Fully-mixed structure reduces the inferred bulk metallicity. The absolute difference can go up to 17\%. For massive planets ($M_{\rm p}>4~M_{\rm J}$), this choice of internal structure, along with its effect on the envelope opacity, has a greater effect than the H-He EOS.
    \item For $M_{\rm p}>0.6~M_{\rm J}$, the observational uncertainties on radii and ages lead to uncertainties in the inferred metallicity (up to 31\%) which are larger than the ones associated with the used H-He EOS and the assumed interior structure. However, for planets with $0.2<M_{\rm p}<0.6~M_{\rm J}$, the theoretical uncertainties are larger.
    \item Using CMS19+HG23 and assuming a Core+envelope structure generally yield lower metallicities for most exoplanets. There are a few exceptions, especially when $M_{\rm p}>4~M_{\rm J}$, for which a higher metallicity can be inferred (due to the planets' position within the temperature-pressure-density space during its evolution, Fig.~\ref{figure:hg23_vs_scvh}).
\end{enumerate}

In this work, we modelled planets with either Core+envelope or Fully-mixed structures. However, in reality, the internal structures of giant planets may be more complex including fuzzy cores and composition gradients. 
As a result, our models should be viewed as the two limiting cases and future work should include more sophisticated internal structures.  
Interestingly, if giant exoplanets have non-homogeneous structures, the atmospheric composition does not represent the planetary bulk composition \citep{knierim2024}. Indeed, \citet{swain2024} suggested that the discrepancy between the 
mass-metallicty relation based on atmospheric measurements and the one inferred from interior models may indicate the presence of composition gradients. 

A better characterisation of giant exoplanets is expected from several fronts. First, improved experimental and theoretical research can refine the EOSs for H-He, heavy elements, and their mixtures as well as phase separations. Second, a deeper understanding of Jupiter and Saturn reveals information on the fundamental properties of gas giant planets. Third, upcoming  space missions  will provide accurate  measurements of the planetary mass, radius and age (Plato) and determine the   atmospheric compositions (Ariel) to further constrain the interiors of exoplanets. 
Overall, combining these different avenues will not only expand our understanding of distant gas giants but also enrich our comprehension of our own planetary system.

\begin{acknowledgements}
We thank Tristan Guillot for insightful discussions. We acknowledge support from SNSF grant \texttt{\detokenize{200020_215634}} and the National Centre for Competence in Research ‘PlanetS’ supported by SNSF.
\end{acknowledgements}

%
   \bibliographystyle{aa} 
   \bibliography{biblio} 

\begin{thebibliography}{49}
\expandafter\ifx\csname natexlab\endcsname\relax\def\natexlab#1{#1}\fi

\bibitem[{{Asplund} {et~al.}(2021){Asplund}, {Amarsi}, \& {Grevesse}}]{asplund2021}
{Asplund}, M., {Amarsi}, A.~M., \& {Grevesse}, N. 2021, \aap, 653, A141

\bibitem[{{Baraffe} {et~al.}(2008){Baraffe}, {Chabrier}, \& {Barman}}]{baraffe2008}
{Baraffe}, I., {Chabrier}, G., \& {Barman}, T. 2008, \aap, 482, 315

\bibitem[{{Bloot} {et~al.}(2023){Bloot}, {Miguel}, {Bazot}, \& {Howard}}]{bloot2023}
{Bloot}, S., {Miguel}, Y., {Bazot}, M., \& {Howard}, S. 2023, \mnras, 523, 6282

\bibitem[{{Bolton} {et~al.}(2017){Bolton}, {Adriani}, {Adumitroaie}, {Allison}, {Anderson}, {Atreya}, {Bloxham}, {Brown}, {Connerney}, {DeJong}, {Folkner}, {Gautier}, {Grassi}, {Gulkis}, {Guillot}, {Hansen}, {Hubbard}, {Iess}, {Ingersoll}, {Janssen}, {Jorgensen}, {Kaspi}, {Levin}, {Li}, {Lunine}, {Miguel}, {Mura}, {Orton}, {Owen}, {Ravine}, {Smith}, {Steffes}, {Stone}, {Stevenson}, {Thorne}, {Waite}, {Durante}, {Ebert}, {Greathouse}, {Hue}, {Parisi}, {Szalay}, \& {Wilson}}]{bolton2017}
{Bolton}, S.~J., {Adriani}, A., {Adumitroaie}, V., {et~al.} 2017, Science, 356, 821

\bibitem[{{Borucki} {et~al.}(2010){Borucki}, {Koch}, {Basri}, {Batalha}, {Brown}, {Caldwell}, {Caldwell}, {Christensen-Dalsgaard}, {Cochran}, {DeVore}, {Dunham}, {Dupree}, {Gautier}, {Geary}, {Gilliland}, {Gould}, {Howell}, {Jenkins}, {Kondo}, {Latham}, {Marcy}, {Meibom}, {Kjeldsen}, {Lissauer}, {Monet}, {Morrison}, {Sasselov}, {Tarter}, {Boss}, {Brownlee}, {Owen}, {Buzasi}, {Charbonneau}, {Doyle}, {Fortney}, {Ford}, {Holman}, {Seager}, {Steffen}, {Welsh}, {Rowe}, {Anderson}, {Buchhave}, {Ciardi}, {Walkowicz}, {Sherry}, {Horch}, {Isaacson}, {Everett}, {Fischer}, {Torres}, {Johnson}, {Endl}, {MacQueen}, {Bryson}, {Dotson}, {Haas}, {Kolodziejczak}, {Van Cleve}, {Chandrasekaran}, {Twicken}, {Quintana}, {Clarke}, {Allen}, {Li}, {Wu}, {Tenenbaum}, {Verner}, {Bruhweiler}, {Barnes}, \& {Prsa}}]{borucki2010}
{Borucki}, W.~J., {Koch}, D., {Basri}, G., {et~al.} 2010, Science, 327, 977

\bibitem[{{Bouchy} {et~al.}(2017){Bouchy}, {Doyon}, {Artigau}, {Melo}, {Hernandez}, {Wildi}, {Delfosse}, {Lovis}, {Figueira}, {Canto Martins}, {Gonz{\'a}lez Hern{\'a}ndez}, {Thibault}, {Reshetov}, {Pepe}, {Santos}, {de Medeiros}, {Rebolo}, {Abreu}, {Adibekyan}, {Bandy}, {Benz}, {Blind}, {Bohlender}, {Boisse}, {Bovay}, {Broeg}, {Brousseau}, {Cabral}, {Chazelas}, {Cloutier}, {Coelho}, {Conod}, {Cumming}, {Delabre}, {Genolet}, {Hagelberg}, {Jayawardhana}, {K{\"a}ufl}, {Lafreni{\`e}re}, {de Castro Le{\~a}o}, {Malo}, {de Medeiros Martins}, {Matthews}, {Metchev}, {Oshagh}, {Ouellet}, {Parro}, {Rasilla Pi{\~n}eiro}, {Santos}, {Sarajlic}, {Segovia}, {Sordet}, {Udry}, {Valencia}, {Vall{\'e}e}, {Venn}, {Wade}, \& {Saddlemyer}}]{bouchy2017}
{Bouchy}, F., {Doyon}, R., {Artigau}, {\'E}., {et~al.} 2017, The Messenger, 169, 21

\bibitem[{{Chabrier} {et~al.}(2019){Chabrier}, {Mazevet}, \& {Soubiran}}]{chabrier2019}
{Chabrier}, G., {Mazevet}, S., \& {Soubiran}, F. 2019, \apj, 872, 51

\bibitem[{{Edwards} {et~al.}(2023){Edwards}, {Changeat}, {Tsiaras}, {Yip}, {Al-Refaie}, {Anisman}, {Bieger}, {Gressier}, {Shibata}, {Skaf}, {Bouwman}, {Cho}, {Ikoma}, {Venot}, {Waldmann}, {Lagage}, \& {Tinetti}}]{edwards2023}
{Edwards}, B., {Changeat}, Q., {Tsiaras}, A., {et~al.} 2023, \apjs, 269, 31

\bibitem[{{Fortney} {et~al.}(2021){Fortney}, {Dawson}, \& {Komacek}}]{fortney2021}
{Fortney}, J.~J., {Dawson}, R.~I., \& {Komacek}, T.~D. 2021, Journal of Geophysical Research (Planets), 126, e06629

\bibitem[{{Gardner} {et~al.}(2006){Gardner}, {Mather}, {Clampin}, {Doyon}, {Greenhouse}, {Hammel}, {Hutchings}, {Jakobsen}, {Lilly}, {Long}, {Lunine}, {McCaughrean}, {Mountain}, {Nella}, {Rieke}, {Rieke}, {Rix}, {Smith}, {Sonneborn}, {Stiavelli}, {Stockman}, {Windhorst}, \& {Wright}}]{gardner2006}
{Gardner}, J.~P., {Mather}, J.~C., {Clampin}, M., {et~al.} 2006, \ssr, 123, 485

\bibitem[{{Guillot}(2005)}]{guillot2005}
{Guillot}, T. 2005, Annual Review of Earth and Planetary Sciences, 33, 493

\bibitem[{{Guillot} \& {Morel}(1995)}]{guillot1995_cepam}
{Guillot}, T. \& {Morel}, P. 1995, \aaps, 109, 109

\bibitem[{{Helled}(2023)}]{helled2023}
{Helled}, R. 2023, \aap, 675, L8

\bibitem[{{Helled} \& {Howard}(2024)}]{helledhoward2024}
{Helled}, R. \& {Howard}, S. 2024, arXiv e-prints, arXiv:2407.05853

\bibitem[{{Helled} \& {Stevenson}(2024)}]{helled2024}
{Helled}, R. \& {Stevenson}, D.~J. 2024, AGU Advances, 5, e2024AV001171

\bibitem[{{Howard} \& {Guillot}(2023)}]{howardguillot2023}
{Howard}, S. \& {Guillot}, T. 2023, \aap, 672, L1

\bibitem[{{Howard} {et~al.}(2023{\natexlab{a}}){Howard}, {Guillot}, {Bazot}, {Miguel}, {Stevenson}, {Galanti}, {Kaspi}, {Hubbard}, {Militzer}, {Helled}, {Nettelmann}, {Idini}, \& {Bolton}}]{howard2023_interior}
{Howard}, S., {Guillot}, T., {Bazot}, M., {et~al.} 2023{\natexlab{a}}, \aap, 672, A33

\bibitem[{{Howard} {et~al.}(2023{\natexlab{b}}){Howard}, {Guillot}, {Markham}, {Helled}, {M{\"u}ller}, {Stevenson}, {Lunine}, {Miguel}, \& {Nettelmann}}]{howard_invZ}
{Howard}, S., {Guillot}, T., {Markham}, S., {et~al.} 2023{\natexlab{b}}, \aap, 680, L2

\bibitem[{{Howard} {et~al.}(2024){Howard}, {M{\"u}ller}, \& {Helled}}]{howard2024}
{Howard}, S., {M{\"u}ller}, S., \& {Helled}, R. 2024, \aap, 689, A15

\bibitem[{{Hubbard} \& {Marley}(1989)}]{hubbard1989}
{Hubbard}, W.~B. \& {Marley}, M.~S. 1989, \icarus, 78, 102

\bibitem[{{Hummer} \& {Mihalas}(1988)}]{hummer1988}
{Hummer}, D.~G. \& {Mihalas}, D. 1988, \apj, 331, 794

\bibitem[{{Knierim} \& {Helled}(2024)}]{knierim2024}
{Knierim}, H. \& {Helled}, R. 2024, arXiv e-prints, arXiv:2407.09341

\bibitem[{{Kreidberg} {et~al.}(2014){Kreidberg}, {Bean}, {D{\'e}sert}, {Line}, {Fortney}, {Madhusudhan}, {Stevenson}, {Showman}, {Charbonneau}, {McCullough}, {Seager}, {Burrows}, {Henry}, {Williamson}, {Kataria}, \& {Homeier}}]{kreidberg2014}
{Kreidberg}, L., {Bean}, J.~L., {D{\'e}sert}, J.-M., {et~al.} 2014, \apjl, 793, L27

\bibitem[{{Lyon} \& {Johnson}(1992)}]{sesame1992}
{Lyon}, S.~P. \& {Johnson}, J.~D. 1992, LANL Report, LA-UR-92-3407

\bibitem[{{Madhusudhan} {et~al.}(2014){Madhusudhan}, {Crouzet}, {McCullough}, {Deming}, \& {Hedges}}]{madhusudhan2014}
{Madhusudhan}, N., {Crouzet}, N., {McCullough}, P.~R., {Deming}, D., \& {Hedges}, C. 2014, \apjl, 791, L9

\bibitem[{{Mankovich} \& {Fuller}(2021)}]{mankovich2021}
{Mankovich}, C.~R. \& {Fuller}, J. 2021, Nature Astronomy, 5, 1103

\bibitem[{{Mayor} {et~al.}(2003){Mayor}, {Pepe}, {Queloz}, {Bouchy}, {Rupprecht}, {Lo Curto}, {Avila}, {Benz}, {Bertaux}, {Bonfils}, {Dall}, {Dekker}, {Delabre}, {Eckert}, {Fleury}, {Gilliotte}, {Gojak}, {Guzman}, {Kohler}, {Lizon}, {Longinotti}, {Lovis}, {Megevand}, {Pasquini}, {Reyes}, {Sivan}, {Sosnowska}, {Soto}, {Udry}, {van Kesteren}, {Weber}, \& {Weilenmann}}]{mayor2003}
{Mayor}, M., {Pepe}, F., {Queloz}, D., {et~al.} 2003, The Messenger, 114, 20

\bibitem[{{Miguel} {et~al.}(2022){Miguel}, {Bazot}, {Guillot}, {Howard}, {Galanti}, {Kaspi}, {Hubbard}, {Militzer}, {Helled}, {Atreya}, {Connerney}, {Durante}, {Kulowski}, {Lunine}, {Stevenson}, \& {Bolton}}]{miguel2022}
{Miguel}, Y., {Bazot}, M., {Guillot}, T., {et~al.} 2022, \aap, 662, A18

\bibitem[{{Militzer} \& {Hubbard}(2013)}]{militzer2013}
{Militzer}, B. \& {Hubbard}, W.~B. 2013, \apj, 774, 148

\bibitem[{{Mordasini} {et~al.}(2016){Mordasini}, {van Boekel}, {Molli{\`e}re}, {Henning}, \& {Benneke}}]{mordasini2016}
{Mordasini}, C., {van Boekel}, R., {Molli{\`e}re}, P., {Henning}, T., \& {Benneke}, B. 2016, \apj, 832, 41

\bibitem[{{M{\"u}ller} {et~al.}(2020){M{\"u}ller}, {Ben-Yami}, \& {Helled}}]{muller2020}
{M{\"u}ller}, S., {Ben-Yami}, M., \& {Helled}, R. 2020, \apj, 903, 147

\bibitem[{{M{\"u}ller} \& {Helled}(2021)}]{muller2021}
{M{\"u}ller}, S. \& {Helled}, R. 2021, \mnras, 507, 2094

\bibitem[{{M{\"u}ller} \& {Helled}(2023)}]{muller2023}
{M{\"u}ller}, S. \& {Helled}, R. 2023, \aap, 669, A24

\bibitem[{{M{\"u}ller} \& {Helled}(2024)}]{muller2024}
{M{\"u}ller}, S. \& {Helled}, R. 2024, \apj, 967, 7

\bibitem[{{Otegi} {et~al.}(2020){Otegi}, {Bouchy}, \& {Helled}}]{otegi2020}
{Otegi}, J.~F., {Bouchy}, F., \& {Helled}, R. 2020, \aap, 634, A43

\bibitem[{{Parc} {et~al.}(2024){Parc}, {Bouchy}, {Venturini}, {Dorn}, \& {Helled}}]{parc2024}
{Parc}, L., {Bouchy}, F., {Venturini}, J., {Dorn}, C., \& {Helled}, R. 2024, \aap, 688, A59

\bibitem[{{Parmentier} {et~al.}(2015){Parmentier}, {Guillot}, {Fortney}, \& {Marley}}]{parmentier2015}
{Parmentier}, V., {Guillot}, T., {Fortney}, J.~J., \& {Marley}, M.~S. 2015, \aap, 574, A35

\bibitem[{{Pepe} {et~al.}(2021){Pepe}, {Cristiani}, {Rebolo}, {Santos}, {Dekker}, {Cabral}, {Di Marcantonio}, {Figueira}, {Lo Curto}, {Lovis}, {Mayor}, {M{\'e}gevand}, {Molaro}, {Riva}, {Zapatero Osorio}, {Amate}, {Manescau}, {Pasquini}, {Zerbi}, {Adibekyan}, {Abreu}, {Affolter}, {Alibert}, {Aliverti}, {Allart}, {Allende Prieto}, {{\'A}lvarez}, {Alves}, {Avila}, {Baldini}, {Bandy}, {Barros}, {Benz}, {Bianco}, {Borsa}, {Bourrier}, {Bouchy}, {Broeg}, {Calderone}, {Cirami}, {Coelho}, {Conconi}, {Coretti}, {Cumani}, {Cupani}, {D'Odorico}, {Damasso}, {Deiries}, {Delabre}, {Demangeon}, {Dumusque}, {Ehrenreich}, {Faria}, {Fragoso}, {Genolet}, {Genoni}, {G{\'e}nova Santos}, {Gonz{\'a}lez Hern{\'a}ndez}, {Hughes}, {Iwert}, {Kerber}, {Knudstrup}, {Landoni}, {Lavie}, {Lillo-Box}, {Lizon}, {Maire}, {Martins}, {Mehner}, {Micela}, {Modigliani}, {Monteiro}, {Monteiro}, {Moschetti}, {Murphy}, {Nunes}, {Oggioni}, {Oliveira}, {Oshagh}, {Pall{\'e}}, {Pariani}, {Poretti}, {Rasilla}, {Rebord{\~a}o}, {Redaelli}, {Santana Tschudi},
  {Santin}, {Santos}, {S{\'e}gransan}, {Schmidt}, {Segovia}, {Sosnowska}, {Sozzetti}, {Sousa}, {Span{\`o}}, {Su{\'a}rez Mascare{\~n}o}, {Tabernero}, {Tenegi}, {Udry}, \& {Zanutta}}]{pepe2021}
{Pepe}, F., {Cristiani}, S., {Rebolo}, R., {et~al.} 2021, \aap, 645, A96

\bibitem[{{Rauer} {et~al.}(2014){Rauer}, {Catala}, {Aerts}, {Appourchaux}, {Benz}, {Brandeker}, {Christensen-Dalsgaard}, {Deleuil}, {Gizon}, {Goupil}, {G{\"u}del}, {Janot-Pacheco}, {Mas-Hesse}, {Pagano}, {Piotto}, {Pollacco}, {Santos}, {Smith}, {Su{\'a}rez}, {Szab{\'o}}, {Udry}, {Adibekyan}, {Alibert}, {Almenara}, {Amaro-Seoane}, {Eiff}, {Asplund}, {Antonello}, {Barnes}, {Baudin}, {Belkacem}, {Bergemann}, {Bihain}, {Birch}, {Bonfils}, {Boisse}, {Bonomo}, {Borsa}, {Brand{\~a}o}, {Brocato}, {Brun}, {Burleigh}, {Burston}, {Cabrera}, {Cassisi}, {Chaplin}, {Charpinet}, {Chiappini}, {Church}, {Csizmadia}, {Cunha}, {Damasso}, {Davies}, {Deeg}, {D{\'\i}az}, {Dreizler}, {Dreyer}, {Eggenberger}, {Ehrenreich}, {Eigm{\"u}ller}, {Erikson}, {Farmer}, {Feltzing}, {de Oliveira Fialho}, {Figueira}, {Forveille}, {Fridlund}, {Garc{\'\i}a}, {Giommi}, {Giuffrida}, {Godolt}, {Gomes da Silva}, {Granzer}, {Grenfell}, {Grotsch-Noels}, {G{\"u}nther}, {Haswell}, {Hatzes}, {H{\'e}brard}, {Hekker}, {Helled}, {Heng}, {Jenkins},
  {Johansen}, {Khodachenko}, {Kislyakova}, {Kley}, {Kolb}, {Krivova}, {Kupka}, {Lammer}, {Lanza}, {Lebreton}, {Magrin}, {Marcos-Arenal}, {Marrese}, {Marques}, {Martins}, {Mathis}, {Mathur}, {Messina}, {Miglio}, {Montalban}, {Montalto}, {Monteiro}, {Moradi}, {Moravveji}, {Mordasini}, {Morel}, {Mortier}, {Nascimbeni}, {Nelson}, {Nielsen}, {Noack}, {Norton}, {Ofir}, {Oshagh}, {Ouazzani}, {P{\'a}pics}, {Parro}, {Petit}, {Plez}, {Poretti}, {Quirrenbach}, {Ragazzoni}, {Raimondo}, {Rainer}, {Reese}, {Redmer}, {Reffert}, {Rojas-Ayala}, {Roxburgh}, {Salmon}, {Santerne}, {Schneider}, {Schou}, {Schuh}, {Schunker}, {Silva-Valio}, {Silvotti}, {Skillen}, {Snellen}, {Sohl}, {Sousa}, {Sozzetti}, {Stello}, {Strassmeier}, {{\v{S}}vanda}, {Szab{\'o}}, {Tkachenko}, {Valencia}, {Van Grootel}, {Vauclair}, {Ventura}, {Wagner}, {Walton}, {Weingrill}, {Werner}, {Wheatley}, \& {Zwintz}}]{rauer2014}
{Rauer}, H., {Catala}, C., {Aerts}, C., {et~al.} 2014, Experimental Astronomy, 38, 249

\bibitem[{{Ricker} {et~al.}(2015){Ricker}, {Winn}, {Vanderspek}, {Latham}, {Bakos}, {Bean}, {Berta-Thompson}, {Brown}, {Buchhave}, {Butler}, {Butler}, {Chaplin}, {Charbonneau}, {Christensen-Dalsgaard}, {Clampin}, {Deming}, {Doty}, {De Lee}, {Dressing}, {Dunham}, {Endl}, {Fressin}, {Ge}, {Henning}, {Holman}, {Howard}, {Ida}, {Jenkins}, {Jernigan}, {Johnson}, {Kaltenegger}, {Kawai}, {Kjeldsen}, {Laughlin}, {Levine}, {Lin}, {Lissauer}, {MacQueen}, {Marcy}, {McCullough}, {Morton}, {Narita}, {Paegert}, {Palle}, {Pepe}, {Pepper}, {Quirrenbach}, {Rinehart}, {Sasselov}, {Sato}, {Seager}, {Sozzetti}, {Stassun}, {Sullivan}, {Szentgyorgyi}, {Torres}, {Udry}, \& {Villasenor}}]{ricker2015}
{Ricker}, G.~R., {Winn}, J.~N., {Vanderspek}, R., {et~al.} 2015, Journal of Astronomical Telescopes, Instruments, and Systems, 1, 014003

\bibitem[{{Saumon} {et~al.}(1995){Saumon}, {Chabrier}, \& {van Horn}}]{saumon1995}
{Saumon}, D., {Chabrier}, G., \& {van Horn}, H.~M. 1995, \apjs, 99, 713

\bibitem[{{Spilker}(2019)}]{spilker2019}
{Spilker}, L. 2019, Science, 364, 1046

\bibitem[{{Swain} {et~al.}(2024){Swain}, {Hasegawa}, {Thorngren}, \& {Roudier}}]{swain2024}
{Swain}, M.~R., {Hasegawa}, Y., {Thorngren}, D.~P., \& {Roudier}, G.~M. 2024, \ssr, 220, 61

\bibitem[{{Thorngren} \& {Fortney}(2018)}]{thorngren2018}
{Thorngren}, D.~P. \& {Fortney}, J.~J. 2018, \aj, 155, 214

\bibitem[{{Thorngren} {et~al.}(2016){Thorngren}, {Fortney}, {Murray-Clay}, \& {Lopez}}]{thorngren2016}
{Thorngren}, D.~P., {Fortney}, J.~J., {Murray-Clay}, R.~A., \& {Lopez}, E.~D. 2016, \apj, 831, 64

\bibitem[{{Tinetti} {et~al.}(2018){Tinetti}, {Drossart}, {Eccleston}, {Hartogh}, {Heske}, {Leconte}, {Micela}, {Ollivier}, {Pilbratt}, {Puig}, {Turrini}, {Vandenbussche}, {Wolkenberg}, {Beaulieu}, {Buchave}, {Ferus}, {Griffin}, {Guedel}, {Justtanont}, {Lagage}, {Machado}, {Malaguti}, {Min}, {N{\o}rgaard-Nielsen}, {Rataj}, {Ray}, {Ribas}, {Swain}, {Szabo}, {Werner}, {Barstow}, {Burleigh}, {Cho}, {Coud{\'e} du Foresto}, {Coustenis}, {Decin}, {Encrenaz}, {Galand}, {Gillon}, {Helled}, {Morales}, {Garc{\'\i}a Mu{\~n}oz}, {Moneti}, {Pagano}, {Pascale}, {Piccioni}, {Pinfield}, {Sarkar}, {Selsis}, {Tennyson}, {Triaud}, {Venot}, {Waldmann}, {Waltham}, {Wright}, {Amiaux}, {Augu{\`e}res}, {Berth{\'e}}, {Bezawada}, {Bishop}, {Bowles}, {Coffey}, {Colom{\'e}}, {Crook}, {Crouzet}, {Da Peppo}, {Sanz}, {Focardi}, {Frericks}, {Hunt}, {Kohley}, {Middleton}, {Morgante}, {Ottensamer}, {Pace}, {Pearson}, {Stamper}, {Symonds}, {Rengel}, {Renotte}, {Ade}, {Affer}, {Alard}, {Allard}, {Altieri}, {Andr{\'e}}, {Arena}, {Argyriou},
  {Aylward}, {Baccani}, {Bakos}, {Banaszkiewicz}, {Barlow}, {Batista}, {Bellucci}, {Benatti}, {Bernardi}, {B{\'e}zard}, {Blecka}, {Bolmont}, {Bonfond}, {Bonito}, {Bonomo}, {Brucato}, {Brun}, {Bryson}, {Bujwan}, {Casewell}, {Charnay}, {Pestellini}, {Chen}, {Ciaravella}, {Claudi}, {Cl{\'e}dassou}, {Damasso}, {Damiano}, {Danielski}, {Deroo}, {Di Giorgio}, {Dominik}, {Doublier}, {Doyle}, {Doyon}, {Drummond}, {Duong}, {Eales}, {Edwards}, {Farina}, {Flaccomio}, {Fletcher}, {Forget}, {Fossey}, {Fr{\"a}nz}, {Fujii}, {Garc{\'\i}a-Piquer}, {Gear}, {Geoffray}, {G{\'e}rard}, {Gesa}, {Gomez}, {Graczyk}, {Griffith}, {Grodent}, {Guarcello}, {Gustin}, {Hamano}, {Hargrave}, {Hello}, {Heng}, {Herrero}, {Hornstrup}, {Hubert}, {Ida}, {Ikoma}, {Iro}, {Irwin}, {Jarchow}, {Jaubert}, {Jones}, {Julien}, {Kameda}, {Kerschbaum}, {Kervella}, {Koskinen}, {Krijger}, {Krupp}, {Lafarga}, {Landini}, {Lellouch}, {Leto}, {Luntzer}, {Rank-L{\"u}ftinger}, {Maggio}, {Maldonado}, {Maillard}, {Mall}, {Marquette}, {Mathis}, {Maxted}, {Matsuo},
  {Medvedev}, {Miguel}, {Minier}, {Morello}, {Mura}, {Narita}, {Nascimbeni}, {Nguyen Tong}, {Noce}, {Oliva}, {Palle}, {Palmer}, {Pancrazzi}, {Papageorgiou}, {Parmentier}, {Perger}, {Petralia}, {Pezzuto}, {Pierrehumbert}, {Pillitteri}, {Piotto}, {Pisano}, {Prisinzano}, {Radioti}, {R{\'e}ess}, {Rezac}, {Rocchetto}, {Rosich}, {Sanna}, {Santerne}, {Savini}, {Scandariato}, {Sicardy}, {Sierra}, {Sindoni}, {Skup}, {Snellen}, {Sobiecki}, {Soret}, {Sozzetti}, {Stiepen}, {Strugarek}, {Taylor}, {Taylor}, {Terenzi}, {Tessenyi}, {Tsiaras}, {Tucker}, {Valencia}, {Vasisht}, {Vazan}, {Vilardell}, {Vinatier}, {Viti}, {Waters}, {Wawer}, {Wawrzaszek}, {Whitworth}, {Yung}, {Yurchenko}, {Zapatero Osorio}, {Zellem}, {Zingales}, \& {Zwart}}]{tinetti2018}
{Tinetti}, G., {Drossart}, P., {Eccleston}, P., {et~al.} 2018, Experimental Astronomy, 46, 135

\bibitem[{{Turrini} {et~al.}(2018){Turrini}, {Miguel}, {Zingales}, {Piccialli}, {Helled}, {Vazan}, {Oliva}, {Sindoni}, {Pani{\'c}}, {Leconte}, {Min}, {Pirani}, {Selsis}, {Coud{\'e} du Foresto}, {Mura}, \& {Wolkenberg}}]{turrini2018}
{Turrini}, D., {Miguel}, Y., {Zingales}, T., {et~al.} 2018, Experimental Astronomy, 46, 45

\bibitem[{Valencia {et~al.}(2013)Valencia, Guillot, Parmentier, \& Freedman}]{valencia1013}
Valencia, D., Guillot, T., Parmentier, V., \& Freedman, R.~S. 2013, The Astrophysical Journal, 775, 10

\bibitem[{{Wahl} {et~al.}(2017){Wahl}, {Hubbard}, {Militzer}, {Guillot}, {Miguel}, {Movshovitz}, {Kaspi}, {Helled}, {Reese}, {Galanti}, {Levin}, {Connerney}, \& {Bolton}}]{wahl2017}
{Wahl}, S.~M., {Hubbard}, W.~B., {Militzer}, B., {et~al.} 2017, \grl, 44, 4649

\end{thebibliography}
%

\begin{appendix} 

\section{Sample of exoplanets}
\label{app:A}

Table~\ref{tab:param} lists the sample of exoplanets studied in Sec.~\ref{sec:planetS}. We used the PlanetS catalog \citep{otegi2020,parc2024} and selected planets with masses from 0.1 to 10~$M_{\rm J}$ which have relative measurement uncertainties on the mass and radius smaller than 10 and 8\%, respectively. We selected planets with $T_{\rm eq}<1000~$K and with an age estimate. In Sec.~\ref{sec:planetS}, we mention that our sample includes 45 planets. We note that our sample initially included 4 additional planets (shown in bold in the table). However, the lower bound of their measured radius could not even be matched with a pure H-He model (for both EOSs). \citet{muller2020} already reported 6 planets from the \citet{thorngren2016} sample with a similar issue. This may warrant further investigation in the future.

\begin{table*}[h]
\centering
\caption{Our sample of exoplanets.}
\begin{tabular}{l l c c c c c}
\hline
\hline
\# & Name & Mass [$M_{\rm Jup}$] & Radius [$R_{\rm Jup}$] & Age [Gyr] & $T_{\rm eq}$ [K] & Stellar met [dex]\\
\hline
1$_{\rm fm}^\ddagger$ & Kepler-75 b & 9.9 $\pm$ 0.5 & 1.03 $\pm$ 0.06 & 6 $\pm$ 3 & 850 & -0.07 $\pm$ 0.15 \\ 
2$_{\rm fm}$ & TOI-2373 b & 9.3 $\pm$ 0.2 & 0.93 $\pm$ 0.02 & 5.9 $\pm$ 1.7 & 860 & 0.3 $\pm$ 0.05\\ 
3 & TOI-2338 b & 5.98 $\pm$ 0.20 & 1 $\pm$ 0.02 & 7 $\pm$ 2 & 799 & 0.22 $\pm$ 0.04\\ 
4 & WASP-162 b & 5.2 $\pm$ 0.2 & 1 $\pm$ 0.05 & $^{\ast}12.97_{-2.35}^{+0.83}$ & 910 & 0.28 $\pm$ 0.13\\ 
5$^\ddagger$ & Kepler-1704 b & 4.16 $\pm$ 0.29 & $1.066_{-0.042}^{+0.044}$ & $7.4_{-1.0}^{+1.5}$ & 253.8 & 0.196 $\pm$ 0.058\\ 
6$^\ddagger$ & TOI-2589 b & 3.5 $\pm$ 0.1 & 1.08 $\pm$ 0.03 & 11 $\pm$ 2 & 592 & 0.12 $\pm$ 0.04\\ 
7 & TOI-5153 b & 3.26 $\pm$ 0.18 & 1.06 $\pm$ 0.04 & 5.4 $\pm$ 1 & 906 & 0.12 $\pm$ 0.08\\ 
8$^\dagger$ & NGTS-20 b & 2.98 $\pm$ 0.16 & 1.07 $\pm$ 0.04 & 4.1 $\pm$ 2.7 & 688 & 0.15 $\pm$ 0.08\\ 
9 & TOI-2180 b & 2.755 $\pm$ 0.087 & $1.01_{-0.019}^{+0.022}$ & $8.1_{-1.3}^{+1.5}$ & 348 & 0.253 $\pm$ 0.057\\ 
10$^\ddagger$ & HATS-76 b & 2.629 $\pm$ 0.089 & 1.079 $\pm$ 0.031 & $4.6_{-4.0}^{+8.7}$ & 939.8 & $0.322_{-0.049}^{+0.065}$ \\ 
11$^\ddagger$ & TOI-4127 b & 2.30 $\pm$ 0.11 & $1.096_{-0.032}^{+0.039}$ & 4.8 $\pm$ 2.1 & 605.1 & 0.14 $\pm$ 0.12\\ 
12 & K2-114 b & 2.01 $\pm$ 0.12 & 0.932 $\pm$ 0.031& $7.2_{-4.5}^{+4.3}$ & 701 & 0.41 $\pm$ 0.037\\ 
13 & TOI-4515 b & 2.005 $\pm$ 0.052 & 1.086 $\pm$ 0.039 & 1.2 $\pm$ 0.2 & 705 & 0.05 $\pm$ 0.03\\ 
14$^\ddagger$ & HAT-P-15 b & 1.946 $\pm$ 0.066 & 1.072 $\pm$ 0.043 & $6.8_{-1.6}^{+2.5}$ & 904 & 0.22 $\pm$ 0.08\\ 
15$^\ddagger$ & TIC237913194 b & 1.942 $\pm$ 0.092 & $1.117_{-0.047}^{+0.054}$& 5.7 $\pm$ 1.7 & 974 & 0.14 $\pm$ 0.05\\ 
16$^\ddagger$ & Kepler-117 c & 1.84 $\pm$ 0.18 & 1.101 $\pm$ 0.035 & 5.3 $\pm$ 1.4 & 704 & -0.04 $\pm$ 0.1\\ 
17 & HATS-74A b & 1.46 $\pm$ 0.14 & 1.032 $\pm$ 0.021 & $^{\ast}11_{-5.1}^{+2.8}$ & 895.1 & $0.514_{-0.021}^{+0.033}$\\ 
\textbf{18} & \textbf{HATS-77 b} & 1.374 $\pm$ 0.1 & 1.165 $\pm$ 0.021 & 12.1 $\pm$ 5 & 828.3 & 0.253 $\pm$ 0.039\\ 
19 & HATS-17 b & 1.338 $\pm$ 0.065 & 0.777 $\pm$ 0.056 & 2.1 $\pm$ 1.3 & 814 & 0.3 $\pm$ 0.03\\ 
20 & TOI-5542 b & 1.32 $\pm$ 0.1 & 1.009 $\pm$ 0.036 & $10.8_{-3.6}^{+2.1}$ & 441 & -0.21 $\pm$ 0.08\\ 
21$^\dagger$ & TOI-2010 b & 1.286 $\pm$ 0.057 & 1.054 $\pm$ 0.027 & $1.9_{-1.3}^{+2.2}$ & 400.2 & 0.168 $\pm$ 0.055\\ 
22 & WASP-130 b & 1.23 $\pm$ 0.04 & 0.89 $\pm$ 0.03 & $2_{-1.6}^{+5.9}$ & 833 & 0.26 $\pm$ 0.1\\ 
\textbf{23} & \textbf{Kepler-87 b} & 1.02 $\pm$ 0.028 & 1.204 $\pm$ 0.049 & 7.5 $\pm$ 0.5 & 478 & -0.17 $\pm$ 0.03\\ 
24 & TOI-1811 b & 0.972 $\pm$ 0.078 & 0.994 $\pm$ 0.025 & $5.9_{-4.0}^{+4.9}$ & 962.2 & 0.306 $\pm$ 0.077\\ 
\textbf{25} & \textbf{K2-140 b} & 0.93 $\pm$ 0.04 & 1.21 $\pm$ 0.09 & $9.8_{-4.6}^{+3.4}$ & 962 & 0.1 $\pm$ 0.1\\ 
26$^\ddagger$ & TOI-1478 b & 0.851 $\pm$ 0.052 & 1.06 $\pm$ 0.04 & $9.1_{-3.9}^{+3.1}$ & 918 & $0.078_{-0.066}^{+0.072}$\\ 
27$^\dagger$ & K2-290 c & 0.774 $\pm$ 0.047 & 1.006 $\pm$ 0.05 & $4_{-0.8}^{+1.6}$ & 676 & -0.06 $\pm$ 0.1\\ 
28 & HAT-P-54 b & 0.76 $\pm$ 0.032 & 0.944 $\pm$ 0.028 & $3.9_{-2.1}^{+4.3}$ & 818 & -0.127 $\pm$ 0.08\\ 
29$^\dagger$ & TOI-3714 b & 0.689 $\pm$ 0.03 & 0.944 $\pm$ 0.02& $^{\ast}12.5_{-6.5}^{+1.3}$ & 764 & 0.39 $\pm$ 0.086\\ 
30 & WASP-84 b & 0.687 $\pm$ 0.033 & 0.976 $\pm$ 0.025 & 2.1 $\pm$ 1.6 & 833 & 0.09 $\pm$ 0.12\\ 
31$^\dagger$ & HAT-P-17 b & 0.534 $\pm$ 0.018 & 1.01 $\pm$ 0.029 & 7.8 $\pm$ 3.3& 792 & 0 $\pm$ 0.08 \\ 
32 & HATS-75 b & 0.491 $\pm$ 0.039 & 0.884 $\pm$ 0.013 & $^{\ast}13.8_{-3.2}^{+0.0}$ & 772.3 & $0.522_{-0.028}^{+0.051}$\\ 
33 & TOI-201 b & 0.42 $\pm$ 0.05 & $1.008_{-0.015}^{+0.012}$ & $0.87_{-0.49}^{+0.46}$ & 759 & 0.24 $\pm$ 0.036\\ 
34 & TOI-5344 b & 0.412 $\pm$ 0.04 & 0.946 $\pm$ 0.021 & $^{\ast}10.9_{-5.8}^{+2.9}$ & 689 & 0.425 $\pm$ 0.088\\ 
35 & WASP-132 b & 0.41 $\pm$ 0.03 & 0.87 $\pm$ 0.03 & $0.5_{-0.0}^{+2.0}$ & 763 & 0.22 $\pm$ 0.13\\ 
\textbf{36} & \textbf{HATS-47 b} & $0.369_{-0.021}^{+0.031}$ & 1.117 $\pm$ 0.014& 8.1 & 852.9 & -0.113 $\pm$ 0.035\\ 
37 & HATS-49 b & $0.353_{-0.027}^{+0.038}$ & 0.765 $\pm$ 0.013 & $10.5_{-2.0}^{+1.4}$ & 834.8 & 0.208 $\pm$ 0.053\\ 
38 & K2-287 b & 0.315 $\pm$ 0.027 & 0.847 $\pm$ 0.013 & 4.7 $\pm$ 1.0 & 804 & 0.2 $\pm$ 0.04\\ 
39 & TOI-1268 b & 0.30331 $\pm$ 0.026 & 0.812 $\pm$ 0.054 & 0.245 $\pm$ 0.135 & 918.9 & 0.36 $\pm$ 0.06\\ 
40$^\dagger$ & TOI-4406 b & 0.3 $\pm$ 0.03 & 1 $\pm$ 0.02 & 2.9 $\pm$ 0.7 & 904 & 0.1 $\pm$ 0.05\\ 
41$^\dagger$ & HAT-P-19 b & 0.277 $\pm$ 0.017 & 1.008 $\pm$ 0.014 & 7.2 $\pm$ 4 & 981.2 & 0.166 $\pm$ 0.07\\ 
42$^\dagger$ & WASP-69 b & 0.26 $\pm$ 0.017 & 1.057 $\pm$ 0.047 & $2_{-1.5}^{+1.0}$ & 963 & 0.144 $\pm$ 0.077\\ 
43 & K2-329 b & 0.26 $\pm$ 0.022 & 0.774 $\pm$ 0.026 & $1.8_{-1.3}^{+2.2}$ & 650 & $0.098_{-0.07}^{+0.065}$\\ 
44 & HD332231 b & 0.244 $\pm$ 0.021 & 0.867 $\pm$ 0.027 & $4.3_{-1.9}^{+2.5}$ & 876 & 0.036 $\pm$ 0.059\\ 
45 & TOI-3629 b & 0.243 $\pm$ 0.02 & 0.74 $\pm$ 0.014 & $^{\ast}9.8_{-4.8}^{+4.0}$ & 711 & 0.549 $\pm$ 0.093\\ 
46 & HAT-P-12 b & 0.211 $\pm$ 0.012 & $0.959_{-0.021}^{+0.029}$ & 2.5 $\pm$ 2.0 & 963 & -0.29 $\pm$ 0.05\\ 
47 & WASP-156 b & 0.128 $\pm$ 0.01 & 0.51 $\pm$ 0.02 & 6.4 $\pm$ 4 & 970 & 0.24 $\pm$ 0.12 \\ 
48 & HATS-72 b & 0.1254 $\pm$ 0.004& 0.722 $\pm$ 0.003 & $12.17_{-0.45}^{+0.24}$ & 739.3 & 0.099 $\pm$ 0.014\\ 
49$_{\rm ce}$ & TOI-4010 d & 0.12003 $\pm$ 0.01 & $0.551_{-0.013}^{+0.012}$  & 6.1 $\pm$ 3.1 & 650 & 0.37 $\pm$ 0.07\\ 
\hline
\end{tabular}
\label{tab:param}
\begin{flushleft}
\end{flushleft}
\tablefoot{$^\ddagger$ indicates planets for which a model fully made of H-He cannot match the upper bound of the measured radius, for both EOSs. $^\dagger$ is similar but when only CMS19+HG23 does not match the upper bound.
$_{\rm fm}$ indicates planets for which solutions have been found only with the Fully-mixed case.
$_{\rm ce}$ indicates planets for which solutions have been found only with the core+envelope case.
$^{\ast}$ indicates that the age was adapted so that the upper bound does not exceed the age of the galaxy. Planets shown in bold have not been considered in our analysis (see text in Appendix~\ref{app:A}).}
\end{table*}

\clearpage
\newpage
\section{Sampling}
\label{app:C}

\begin{figure}[h]
   \centering
   \includegraphics[width=0.8\hsize]{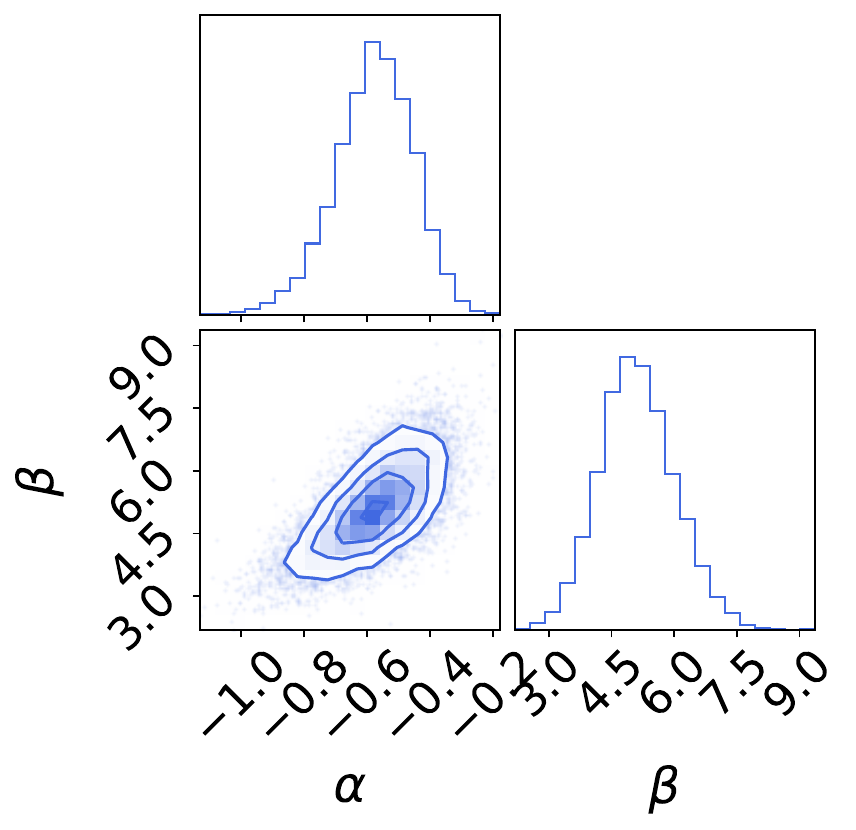}
      \caption{Posterior distributions of the parameters $\alpha$ and $\beta$ in the Core+envelope case.}
         \label{figure:CE_post}
\end{figure}

\begin{figure}[h]
   \centering
   \includegraphics[width=0.8\hsize]{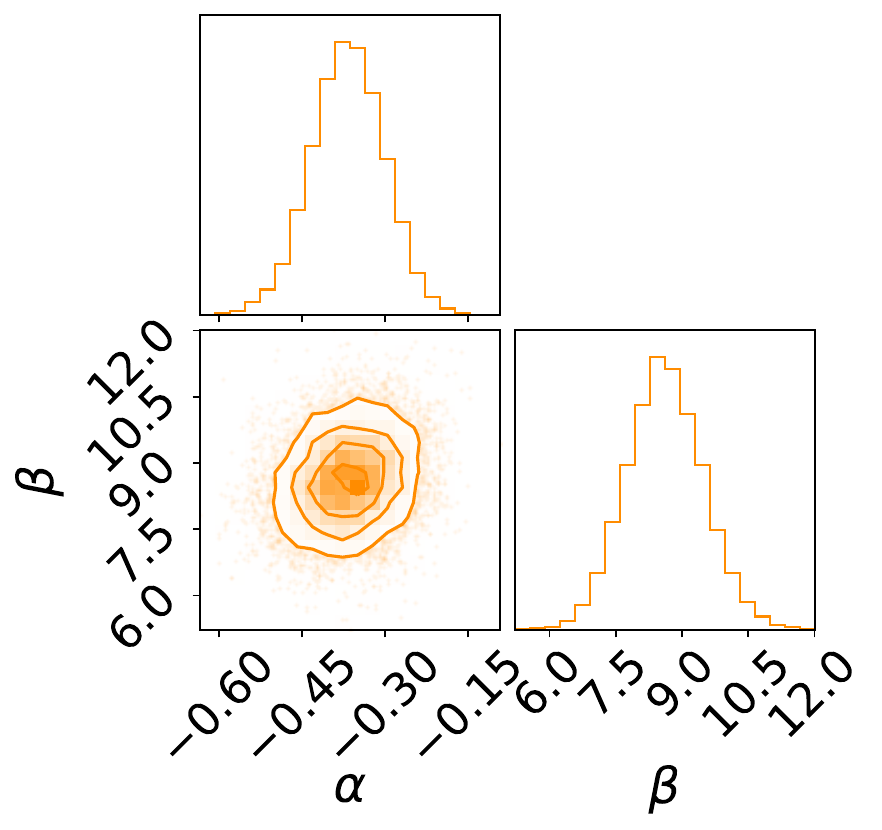}
      \caption{Posterior distributions of the parameters $\alpha$ and $\beta$ in the Fully-mixed case.}
         \label{figure:FM_post}
\end{figure}

Figures~\ref{figure:CE_post} and~\ref{figure:FM_post} show the posteriors distributions of the parameters $\alpha$ and $\beta$ of the power-law $Z_{\rm planet}/Z_{\rm star}=\beta \times M[M_{\rm J}]^{\alpha}$, derived in Sec.~\ref{sec:planetS}. 

\clearpage
\newpage
\section{The mass-metallicity relation}
\label{app:D}

\begin{figure}[h]
   \centering
   \includegraphics[width=\hsize]{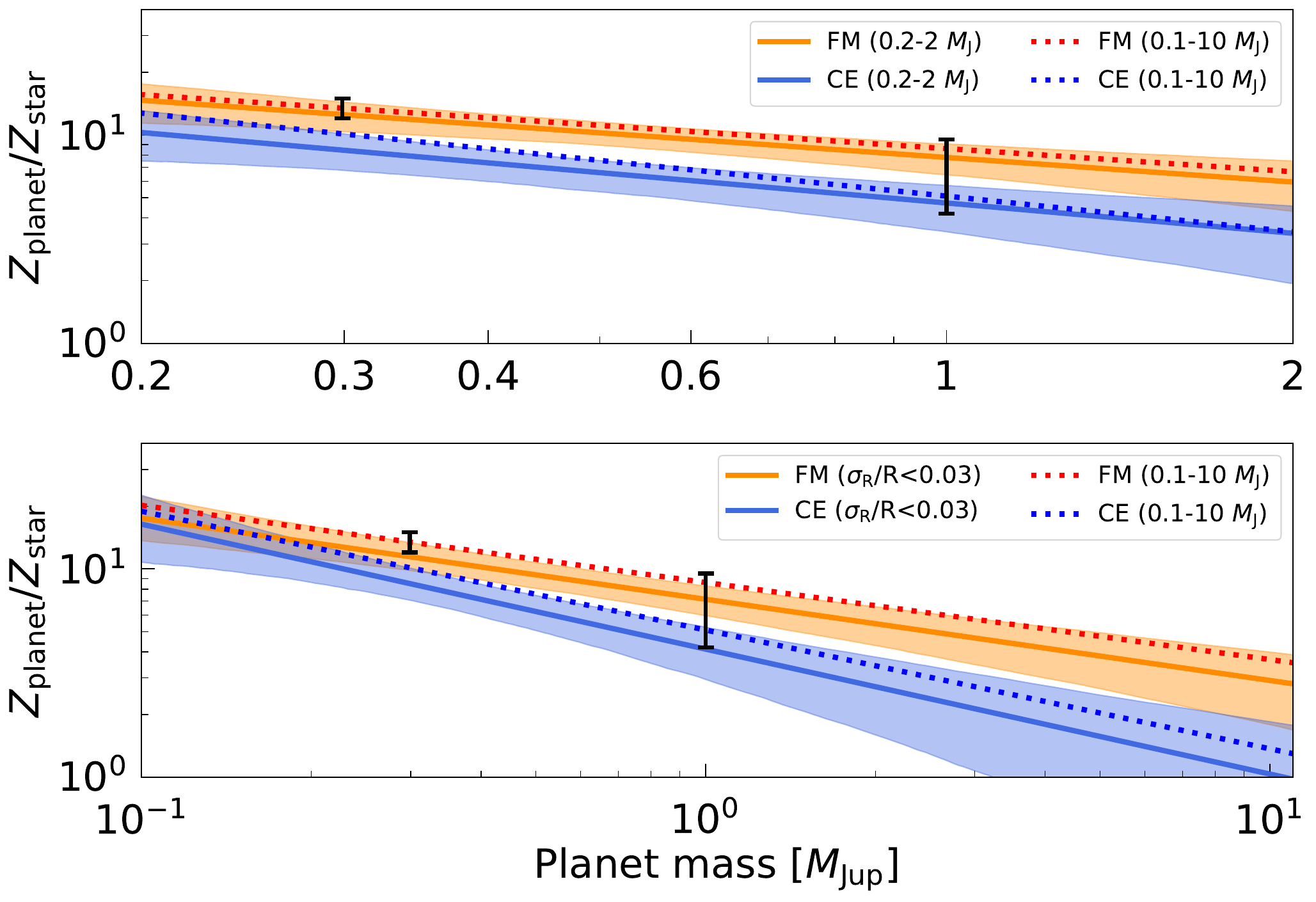}
      \caption{\textit{Top panel}: Mass-metallicity relation considering different planetary mass ranges.  
      Orange and blue solid lines show the best fit for the Fully-mixed (FM) and Core+envelope (CE) cases when considering planets of masses between only 0.2 and 2~$M_{\rm J}$. The shaded areas show the $1\sigma$ error contour around the best fit. The relations previously obtained (see Sec.~\ref{sec:planetS}) when considering the full range of planetary masses (from 0.1 to 10~$M_{\rm J}$) are shown with dotted lines. The metallicities of Jupiter and Saturn \citep{helledhoward2024} are shown in black. \textit{Bottom panel}: Mass-metallicity relation considering different precision on observed radius. 
      Orange and blue solid lines show the best fit for the Fully-mixed (FM) and Core+envelope (CE) cases when only considering planets with relative uncertainties on radii which are less than 3\%. The previously obtained relations, shown with dotted lines, include planets with an uncertainty on radius of up to 8\%.
      }
         \label{figure:fm_vs_ce_app}
\end{figure}

Figure~\ref{figure:fm_vs_ce_app} shows the mass-metallicity relations obtained when considering a different planetary mass range and well as a smaller sample with more accurate measurements. In Sec.~\ref{sec:planetS}, we presented the mass-metallicity relation obtained for the Core+envelope (CE) and Fully-mixed (FM) structures, using CMS19+HG23 and the full range of planetary masses (from 0.1 to 10~$M_{\rm J}$). We found the following values for the parameters $\alpha$ and $\beta$ of the power-law $Z_{\rm planet}/Z_{\rm star}=\beta \times M[M_{\rm J}]^{\alpha}$: $\alpha=-0.37 \pm 0.07$ and $\beta=8.61 \pm 0.88$ in the FM case and $\alpha=-0.57 \pm 0.13$ and $\beta=5.09 \pm 0.95$ in the CE case.

First we use a tighter range of planetary masses that correspond to the giant planet regime, with masses between 0.2 and 2~$M_{\rm J}$ \citep{helled2023}. For this range we now find: $\alpha=-0.379\pm 0.16$ and $\beta=7.81 \pm 1.28$ in the FM case and $\alpha=-0.48 \pm 0.24$ and $\beta=4.72 \pm 1.21$ in the CE case. While the slope of the mass-metallicity relation has not changed in the FM case, it has decreased in the CE case. Overall, with both internal structures, the inferred planetary bulk metallicities are lower than those inferred when considering the full range of planetary masses.

Second, we derived the mass-metallicity relation for the larger mass range as presented in the main text, but considering only planets with a relative error in their observed radius smaller than 3\%. 24 planets from our sample meet this criterion. This time, we find: $\alpha=-0.39\pm 0.11$ and $\beta=7.15 \pm 1.18$ in the FM case and $\alpha=-0.60 \pm 0.23$ and $\beta=4.12 \pm 1.24$ in the CE case. For both internal structures, the slope of these relations remained similar. However, the inferred bulk metallicities are lower than those obtained when considering also planets with larger measurement uncertainties in the planetary radius ($\sigma_{\rm R}/R<8\%$). 
\par 

The analysis presented in this appendix demonstrates the importance of the chosen sample in inferring the mass-metallicity relation. As a result, caution should be taken when comparing relations inferred by different studies. In addition, it is also clear that having more planets with small measurement uncertainties is crucial for the investigation of trends in exoplanetary data.

\end{appendix}

\end{document}